\def\papertitle{Compiling Differentiable Audio Graphs to Real-Time DSP}
\def\paperauthorA{Facundo Franchino}
\def\paperauthorB{Sebastian J. Schlecht}
\documentclass[twoside,a4paper]{article}
\usepackage{etoolbox}

\usepackage{dafx26demo}

\usepackage{amsmath,amssymb,amsfonts,amsthm}
\usepackage{siunitx}
\usepackage{euscript}
\usepackage[T1]{fontenc}
\usepackage[utf8]{inputenc}
\usepackage{ifpdf}
\usepackage[english]{babel}
\usepackage{color}
\usepackage{booktabs}

\input glyphtounicode
\ninept

\newcounter{numauth}
\newcounter{listcnt}
\newcommand\authcnt[1]{\ifdefined#1 \stepcounter{numauth} \fi}

\newcommand\addauth[1]{
\ifdefined#1
\stepcounter{listcnt}
\ifnum \value{listcnt}<\value{numauth}
\appto\authorslist{, #1}
\else
\appto\authorslist{~and~#1}
\fi
\fi}
\authcnt{\paperauthorB}
\authcnt{\paperauthorC}
\authcnt{\paperauthorD}
\def\authorslist{\paperauthorA}
\addauth{\paperauthorB}
\addauth{\paperauthorC}
\addauth{\paperauthorD}
\usepackage{times}

\newif\ifpdf
\ifx\pdfoutput\relax
\else
   \ifcase\pdfoutput
      \pdffalse
   \else
      \pdftrue
   \fi
\fi

\ifpdf
  \usepackage[pdftex,
  pdftitle={\papertitle},
  pdfauthor={\authorslist},
  pdfsubject={Proceedings of the 29th International Conference on Digital Audio Effects (DAFx26)},
  hidelinks,
  bookmarksnumbered,
  pdfstartview=XYZ
]{hyperref}
  \usepackage[pdftex]{graphicx}
\else
  \usepackage[dvips,
  pdftitle={\papertitle},
  pdfauthor={\authorslist},
  pdfsubject={Proceedings of the 29th International Conference on Digital Audio Effects (DAFx26)},
  hidelinks,
  bookmarksnumbered,
  pdfstartview=XYZ
]{hyperref}
\fi
\usepackage[hypcap=true]{caption}
\usepackage{tikz}
\usetikzlibrary{calc}
\usepackage[capitalise]{cleveref}
\crefname{figure}{Fig.}{Figs.}
\Crefname{figure}{Fig.}{Figs.}
\crefname{equation}{Eq.}{Eqs.}
\Crefname{equation}{Eq.}{Eqs.}
\crefname{section}{Sec.}{Secs.}
\Crefname{section}{Sec.}{Secs.}
\title{\papertitle}

\twoaffiliations
{\paperauthorA\sthanks{Corresponding author.}}
{Massachusetts Institute of Technology \\
Cambridge, MA\\
{\tt \href{mailto:facundof@mit.edu}{facundof@mit.edu}}
}
{\paperauthorB}
{Artificial Audio, Multimedia Communications and Signal Processing \\
Friedrich-Alexander-Universit\"at Erlangen-N\"urnberg \\
Erlangen, Germany\\
{\tt \href{mailto:sebastian.schlecht@fau.de}{sebastian.schlecht@fau.de}}
}
\begin{document}
\ifpdf
  \DeclareGraphicsExtensions{.png,.jpg,.pdf}
\else
  \DeclareGraphicsExtensions{.eps}
\fi

\maketitle

\begin{abstract}
Differentiable audio processors are habitually designed and optimised
in machine-learning frameworks, but deploying them as real-time audio
effects still often requires non-automatic implementation in a
dedicated digital signal processing language. The translation is
error-prone, demands an onerous verification process, and detaches
research prototypes from usable production tools. That being so, we
present ADAC, a compiler that lowers a trained model to a
framework-agnostic intermediate representation and emits efficient
FAUST code whose impulse response matches the source model to within
floating-point arithmetic noise, direct paths included. The
optimisation loop is made audible by replacing the model in a running
plugin after each gradient step. The exported processor carries a
small set of macro-controls that leave its stability intact. A
stability certificate computed from the shipped parameters is checked
before the plugin is built. At the demonstration, a feedback
delay network is trained and exported to a working plugin.
\end{abstract}

\vspace{-4mm}

\section{Introduction}
\label{sec:intro}

Differentiable audio frameworks such as FLAMO~\cite{flamo} permit the
optimisation of audio processors by gradient descent. The optimised
model encodes a complete processor, yet it remains confined to its
training framework. FAUST produces efficient real-time code for various targets through a
maintained compilation pathway~\cite{faust2clap}, but a trained model
may only reach FAUST by hand, through a rewriting in said dedicated DSP
language that must be repeated whenever the model changes. Comparable
deployment difficulties come to light in differentiable audio systems such as DDSP~\cite{ddsp}.

\begin{figure*}[!t]
\centering
\resizebox{\textwidth}{!}{%
\begin{tikzpicture}[
    >=latex,
    box/.style={
        draw,
        rectangle,
        align=left,
        inner sep=6pt,
        font=\small\rmfamily
    },
    smallbox/.style={
        draw,
        rectangle,
        align=center,
        inner sep=4pt,
        minimum width=1cm,
        font=\footnotesize\rmfamily
    }
]
%diff model
\node[box] (model) at (0, 0) {
    \textsc{Differentiable Model}\\
    (PyTorch)\\[1ex]
    \centering
    $\bullet \xrightarrow{\nabla} \bullet \xrightarrow{\nabla} \bullet$
};
%json intermediate rep
\node[box] (ir) at (5.5, 0) {
    \textsc{JSON Intermediate Representation}\\[1ex]
    \begin{tabular}{@{}l@{}}
    \ttfamily \{ \\
    \ttfamily \ \ "type": "Recursion",\\
    \ttfamily \ \ "fF": \{ \dots \}\\
    \ttfamily \}
    \end{tabular}
};
%faust dsp
\node[box] (dsp) at (11.5, 0) {
    \textsc{FAUST .dsp}\\[1ex]
    \begin{tabular}{@{}l@{}}
    \ttfamily import("stdfaust.lib");\\
    \ttfamily process = \dots \\
    \ttfamily \ \ par(i, 4, \dots);
    \end{tabular}
};
\draw[->] (model) -- (ir);
\draw[->] (ir) -- (dsp);
%target families inherited from the FAUST backend matrix
\node[smallbox] (plug) at (9.0, -2.3) {Plugins};
\node[smallbox] (web)  at (10.8, -2.3) {Web};
\node[smallbox] (emb)  at (12.6, -2.3) {Embedded};
\node[smallbox] (fpga) at (14.4, -2.3) {FPGA};
\coordinate (bus) at (11.5, -1.5);
\draw (dsp.south) -- (bus);
\draw[->] (bus) -| (plug.north);
\draw[->] (bus) -| (web.north);
\draw[->] (bus) -| (emb.north);
\draw[->] (bus) -| (fpga.north);
\end{tikzpicture}%
}
\caption{Pipeline overview. The differentiable audio graph is extracted from the host framework (PyTorch) into a framework-agnostic JSON intermediate representation, lowered into functional FAUST DSP code, and from there to any FAUST target: audio plugins (VST3, AU, CLAP), the web (WebAssembly), embedded boards (Bela, Daisy, ESP32), and FPGA (VHDL).}
\label{fig:pipeline}
\end{figure*}
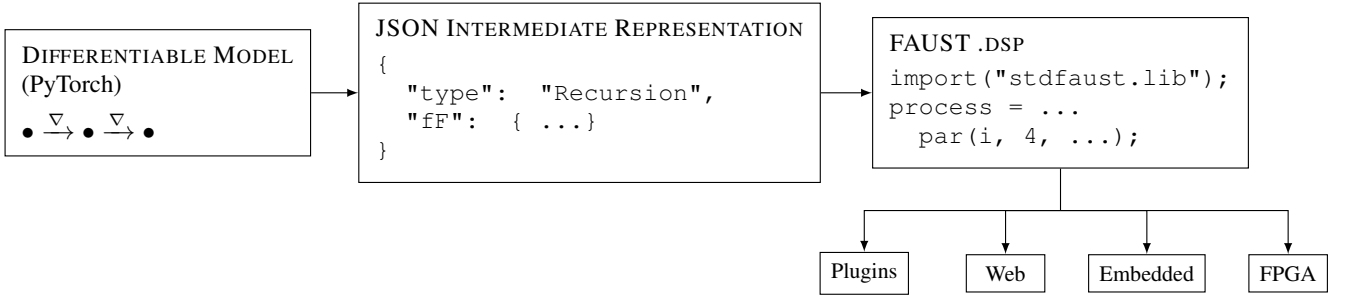

We close this gap with the Automatic Differentiable Audio
Compiler (ADAC). The pipeline (\cref{fig:pipeline}) traverses a trained model's computational
graph, extracts the learned parameters, lowers the processor to a
JSON intermediate representation, and emits equivalent
FAUST~\cite{faust} code, which existing toolchains compile to a
variety of hardware and software targets. This representation captures topology (series, parallel, and
recursive composition) together with numerical parameters, so source
traversal and target emission remain decoupled.

This demonstration uses feedback delay networks
(FDNs)~\cite{jot, schlecht_fdn} as the case study, since they
exercise every composition operator and carry the richest parameter
structure; the same path also compiles a scattering delay
network~\cite{schlecht_scattering} (\cref{sec:conclusion}). The contribution is not solely code generation, but the workflow around it, in three parts. (i)~Training is made audible, the
model replaced in a running plugin after every gradient step.
(ii)~The macro-controls hold the processor stable at any setting.
(iii)~The exported plugin carries a stability certificate, verified
before it is built.

\section{The compiler}
\label{sec:compiler}

\subsection{From model to representation}
\label{ssec:extraction}

An $N$-line FDN is defined by a delay vector
$\mathbf{m} \in \mathbb{N}^N$, a feedback matrix
$\mathbf{A} \in \mathbb{R}^{N \times N}$, input and output gains
$\mathbf{B} \in \mathbb{R}^{N \times N_{\mathrm{in}}}$,
$\mathbf{C} \in \mathbb{R}^{N_{\mathrm{out}} \times N}$, and a direct
path $D \in \mathbb{R}^{N_{\mathrm{out}} \times N_{\mathrm{in}}}$, with
transfer function
\begin{equation}\label{eq:transfer}
  H(z)
  = \mathbf{C}
    \bigl(\boldsymbol{\Delta}(z)^{-1} - \mathbf{A}\bigr)^{\!-1}
    \mathbf{B}
    + D\,,
\end{equation}
where
$\boldsymbol{\Delta}(z)
 = \operatorname{diag}(z^{-m_1}, \ldots, z^{-m_N})$~\cite{schlecht_fdn},
so that the shortest path through the recursive branch traverses at least
$\min_i m_i$ samples.

The extractor walks the model tree and serialises each node into the
representation (\cref{fig:fdn_tree}). Many trained modules hold their
parameters in a parametrisation space rather than an audio space. An
orthogonal feedback matrix~\cite{rocchesso} is stored as skew-symmetric weights and
exponentiated on every forward pass, and a Householder matrix as a
single vector $\mathbf{u}$ from which
$\mathbf{I} - 2\mathbf{u}\mathbf{u}^{\!\top}$ is reconstructed. The
extractor accordingly serialises the \emph{effective} parameter, that
is, the value the model applies to the signal, whilst retaining the
raw weights so that the source model may be reconstructed from the
representation without loss.

\begin{center}
\begin{minipage}{\columnwidth}
\centering
\begin{minipage}[t]{0.58\linewidth}
\vspace{0pt}
\resizebox{\linewidth}{!}{
\begin{tikzpicture}[
    >=latex,
    thick,
    box/.style={
        draw,
        rectangle,
        minimum height=0.5cm,
        minimum width=1.0cm,
        font=\ttfamily\small,
        inner sep=3pt
    },
    edge from parent/.style={
        draw,
        ->,
        thick
    },
    lbl/.style={
        draw=none,
        font=\scriptsize\rmfamily,
        fill=white,
        inner sep=1.5pt
    },
    level 1/.style={sibling distance=3.5cm, level distance=1.0cm},
    level 2/.style={sibling distance=4.2cm, level distance=1.2cm},
    level 3/.style={sibling distance=1.7cm, level distance=1.4cm},
    level 4/.style={sibling distance=1.8cm, level distance=1.4cm},
    level 5/.style={sibling distance=1.8cm, level distance=1.4cm}
]
\node[box] {Shell}
    child {
        node[box] {Parallel}
        child {
            node[box] {Series}
            child {
                node[box] {B}
                edge from parent node[lbl, left] {1$\to$4}
            }
            child {
                node[box] {Recursion}
                child {
                    node[box] {fF}
                    child {
                        node[box, font=\rmfamily\scriptsize] {Delays $\to$ Filters}
                        edge from parent node[lbl, left] {4$\to$4}
                    }
                }
                child {
                    node[box] {fB}
                    child {
                        node[box] {A}
                        edge from parent node[lbl, right] {4$\to$4}
                    }
                }
                edge from parent node[lbl, right] {4$\to$4}
            }
            child {
                node[box] {C}
                edge from parent node[lbl, right] {4$\to$1}
            }
        }
        child {
            node[box] {D}
            edge from parent node[lbl, right] {1$\to$1}
        }
    };
\end{tikzpicture}
}
\end{minipage}\hfill
\begin{minipage}[t]{0.38\linewidth}
\vspace{0pt}
\scriptsize
\resizebox{\linewidth}{!}{%
\fbox{%
\begin{tabular}{@{}l@{}}
\ttfamily \{ \\
\ttfamily \ \ "type": "Parallel",\\
\ttfamily \ \ "nodes": [\\
\ttfamily \ \ \ \ \{ \\
\ttfamily \ \ \ \ \ \ "type": "Series", \\
\ttfamily \ \ \ \ \ \ "nodes": [...] \\
\ttfamily \ \ \ \ \},\\
\ttfamily \ \ \ \ \{ \\
\ttfamily \ \ \ \ \ \ "type": "D",\\
\ttfamily \ \ \ \ \ \ "matrix": [...] \\
\ttfamily \ \ \ \ \}\\
\ttfamily \ \ ]\\
\ttfamily \}
\end{tabular}%
}%
}
\end{minipage}
\captionsetup{hypcap=false}

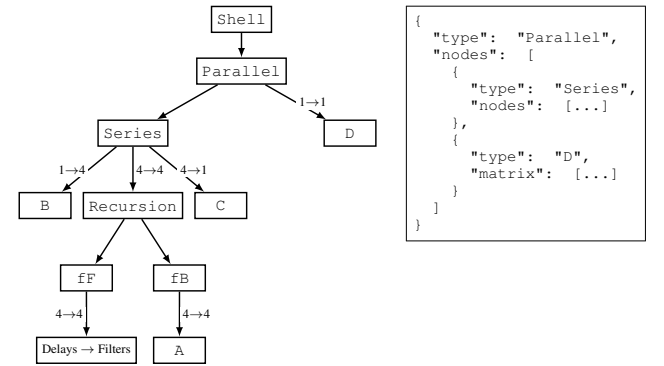
\captionof{figure}{FDN topology. Left: The representation visualised as a tree mapping to state-space formulation matrices. Right: A structural snippet of the corresponding JSON intermediate representation.}
\label{fig:fdn_tree}
\end{minipage}
\end{center}

\subsection{From representation to FAUST}
\label{ssec:codegen}

Each structural node maps to a FAUST composition operator, with
Series becoming \texttt{:}, Parallel a shared-input split, and
Recursion \texttt{\textasciitilde}. Leaves emit the corresponding
primitives, \texttt{@($n$)} for delays, \texttt{*($g$)} for gains,
and \texttt{fi.tf2} cascades for filters, whilst mixing matrices are
hoisted as named functions with explicit sum-of-products
arithmetic. FAUST's recursion operator introduces one
sample of implicit delay. The emitter therefore writes
$z^{-(m_i-1)}$ for each in-loop line, so that with this implicit delay
the round-trip is
\begin{equation}\label{eq:recursion}
  z^{-1}\, z^{-(m_i-1)} = z^{-m_i}\,,
\end{equation}
the exact loop period, and a single $z^{-1}$ on the recursion output,
outside the loop, restores the absolute arrival time that the in-loop
shortening would otherwise advance. When the model's delays are already integer
samples, as in our evaluation, the emitted graph matches the source
model sample-for-sample up to the single-precision arithmetic of the
audio path. Fractional delays, when present, are rounded to the
nearest sample, an error of at most $1/(2f_{\mathrm{s}})$ seconds per
line, where $f_{\mathrm{s}}$ is the sample rate.

\subsection{Correctness}
\label{ssec:correctness}

\begin{figure}[t]
\centerline{\includegraphics[width=\columnwidth]{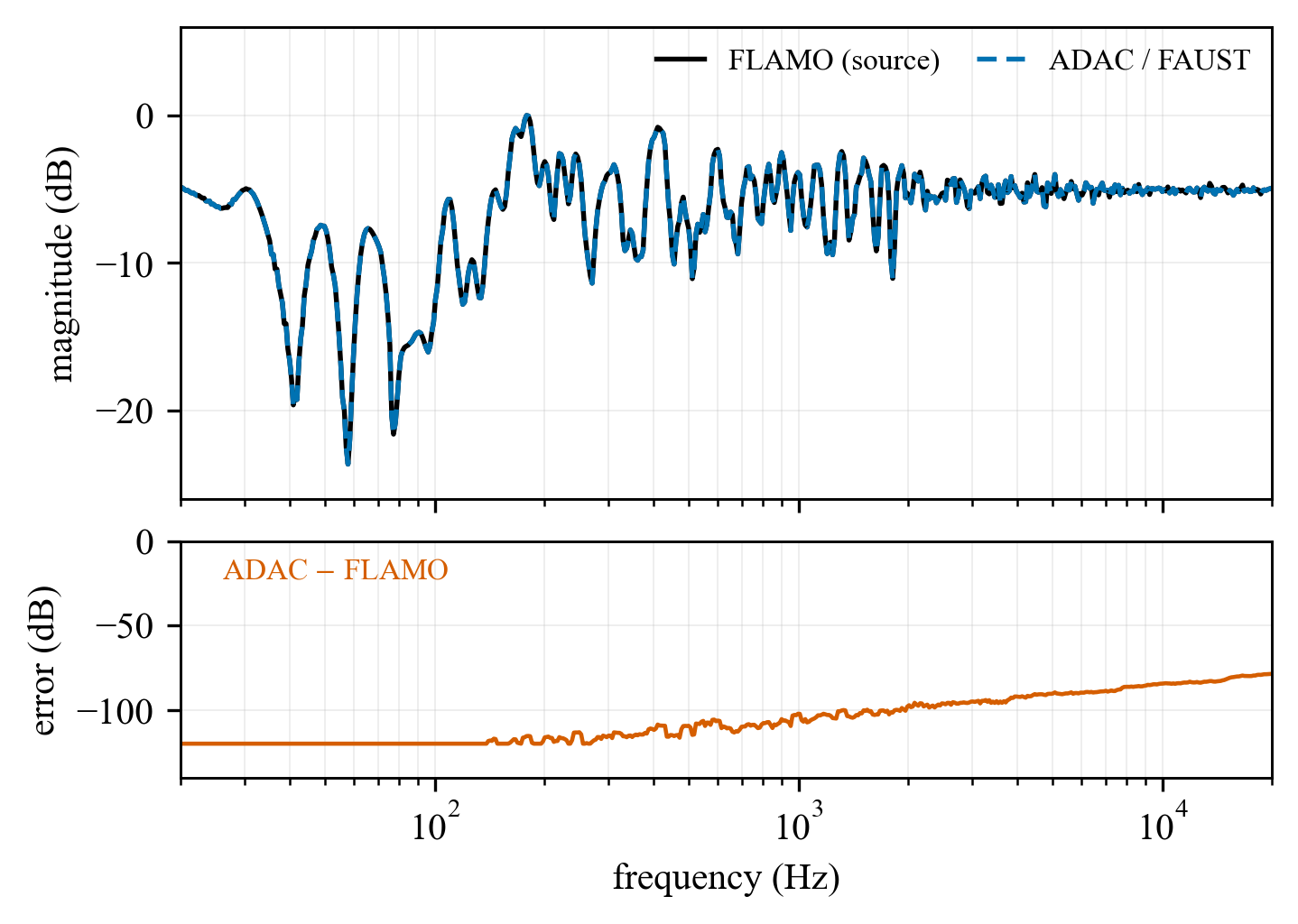}}
\caption{\label{fig:irmatch}{Magnitude response of a compiled four-line FDN, one input-to-output path. Top, the FLAMO source (black) and the emitted FAUST (blue, dashed), $1/12$-octave smoothed; the two coincide across the audible band. Bottom, their difference, at or below $-80$\,\si{\decibel} of the peak, the level of single-precision arithmetic.}}
\end{figure}

The feedback matrix $\mathbf{A}$ is the weighted adjacency matrix of
a directed graph over the delay lines, and entry $(i,j)$ of
$\mathbf{A}^k$ sums the weight products over all length-$k$ walks
between them~\cite{godsil2001}. Equivalently, the FDN is a signal
flow graph whose transfer function is a sum over its paths in the
sense of Mason~\cite{mason1953}. Expanding~\cref{eq:transfer} as a
Neumann series exhibits $H(z)$ as a sum over such walks, so $H(z)$ is
determined entirely by
$(\mathbf{A}, \mathbf{m}, \mathbf{B}, \mathbf{C}, D)$. The pipeline
preserves each of these quantities, and therefore the transfer
function. Across the full input-to-output matrix of a stereo FDN, and
for a mono FDN with a direct path, the impulse responses of the
compiled FAUST agree with the source model to within
$7 \times 10^{-5}$ of peak with no alignment applied
(\cref{fig:irmatch}), a residual at the level of single-precision
arithmetic rather than structural error. The implementation is
exercised by 201 unit tests and an integration suite comparing
impulse responses end-to-end.

\subsection{Cost}
\label{ssec:cost}

\begin{figure}[t]
\centerline{\includegraphics[width=\columnwidth]{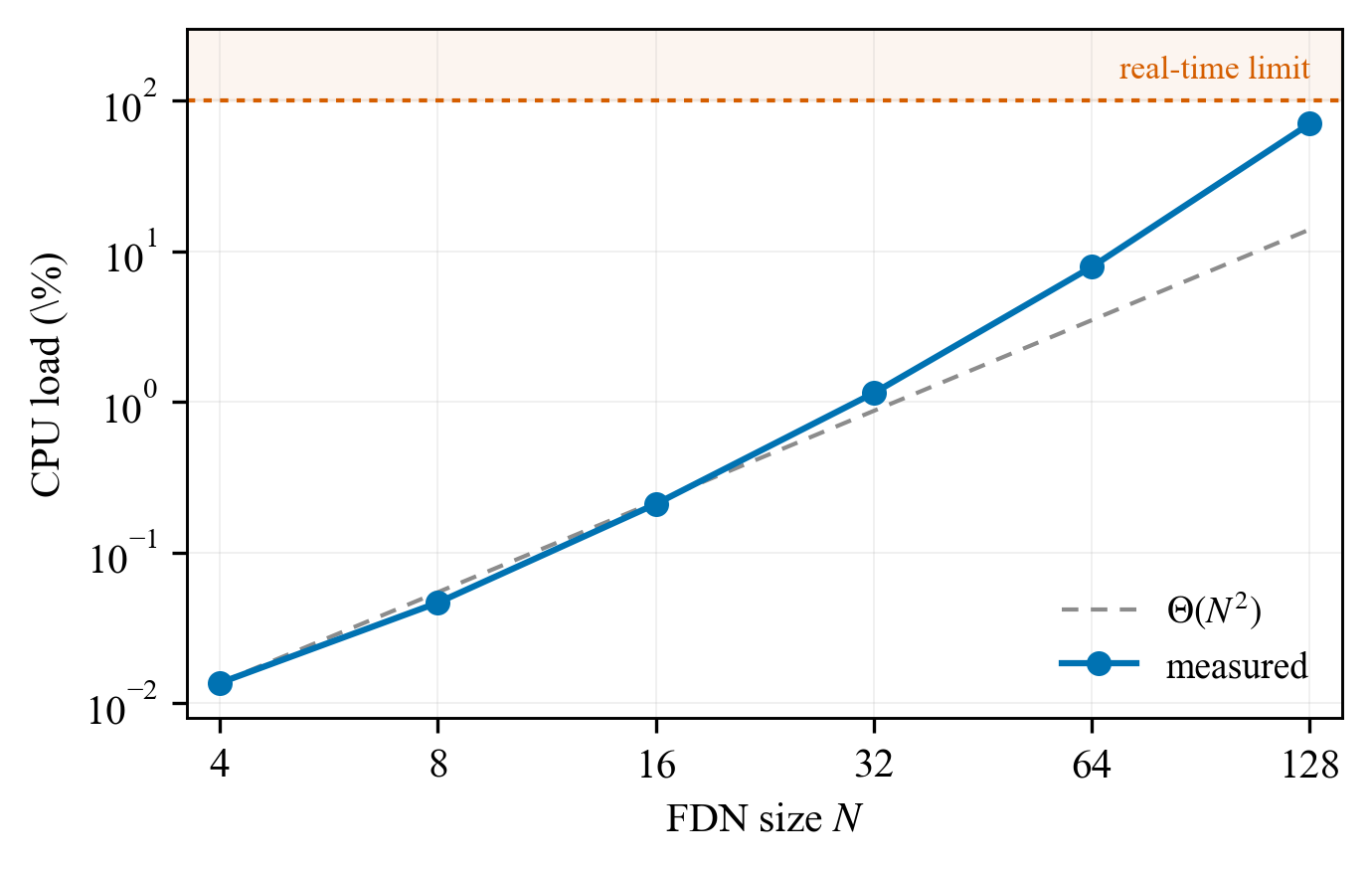}}
\caption{\label{fig:scaling}{Single-core CPU load of the emitted FAUST against FDN size $N$, on an Apple M2 at \SI{48}{\kilo\hertz}. The measured cost tracks the $\Theta(N^2)$ reference until the feedback matrix exceeds the cache. The shaded region marks loads above real time.}}
\end{figure}

For a dense $N$-line FDN with $S$ second-order sections per line, the
emitted code costs $\Theta(N^2 + NS + N_{\mathrm{in}}N
+ N_{\mathrm{out}}N)$ arithmetic operations per sample and
$\Theta(\sum_i m_i + NS)$ state; the matrix emitter drops zero
entries, so the $N^2$ term falls to $\operatorname{nnz}(\mathbf{A})$
for sparse feedback. Benchmarked with \texttt{faust2bench} on a single
core of an Apple M2 at \SI{48}{\kilo\hertz}
(\cref{fig:scaling}), the measured load follows this $\Theta(N^2)$
prediction in the practical range and rises above it once the feedback
matrix outgrows the cache. A 32-line FDN, representative of a practical reverberator, runs
at roughly ninety times real time, and a 64-line one at fourteen, so
even large networks leave ample headroom on commodity hardware.

\section{Macro-controls}
\label{sec:controls}

\begin{figure}[t]
\centerline{\includegraphics[width=\columnwidth]{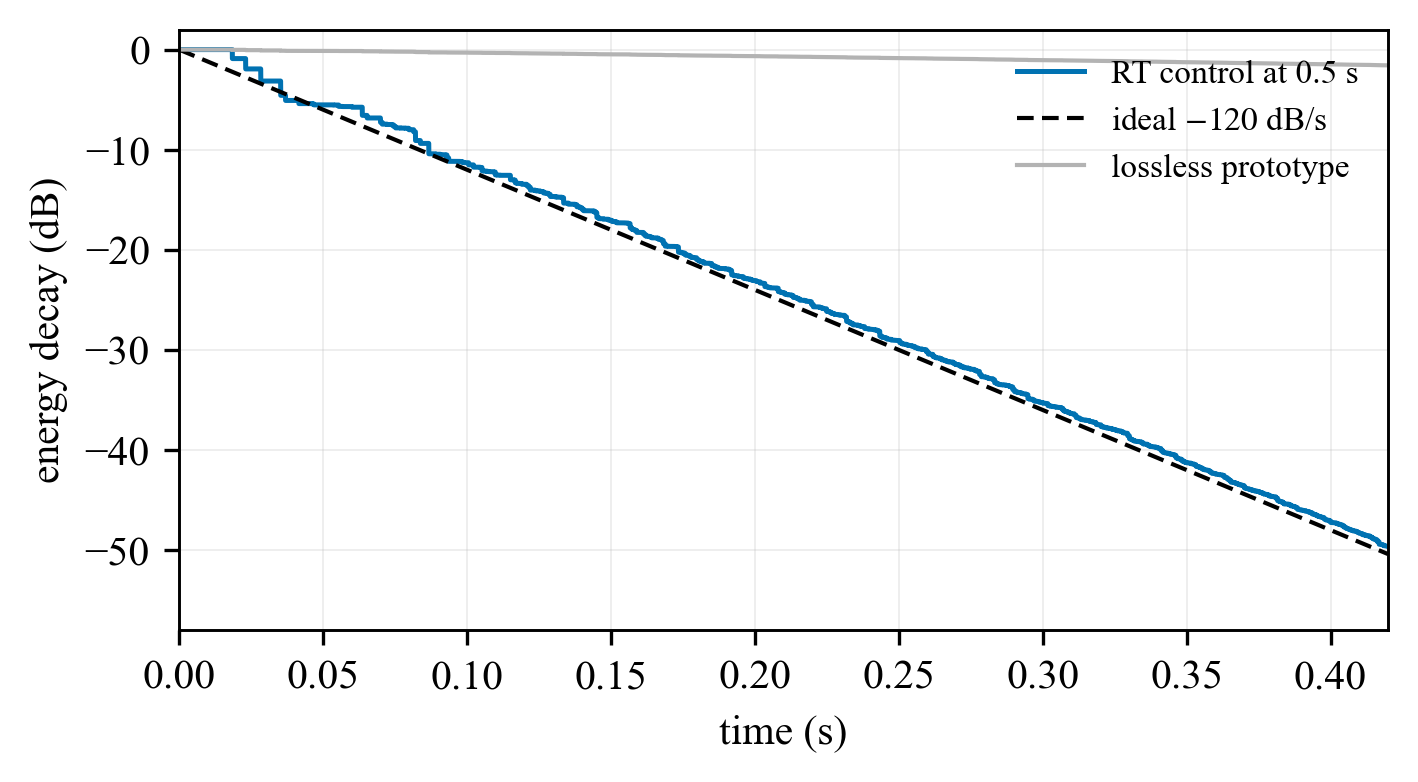}}
\caption{\label{fig:rt60}{Compiled-plugin energy decay for \SI{0.5}{\second} reverberation time, compared with the ideal \SI{-120}{\decibel\per\second} line and lossless prototype.}}
\end{figure}

The raw trained parameters are unsuitable as user controls, being
jointly optimised and resident in parametrisation spaces, so that
moving any one of them in isolation can destabilise the
processor. The compiler instead offers a fixed vocabulary of macro-controls layered
onto the generated code, namely reverberation time, dry/wet balance,
and pre-delay. The reverberation-time control follows
Jot's~\cite{jot} homogeneous-decay construction. Writing
$\mathrm{RT}$ for the reverberation time, the interval over which the
energy decays by \SI{60}{\decibel}, each delay line of length $m_i$
receives the gain
\begin{equation}\label{eq:jot}
  g_i = 10^{-3 m_i / (f_{\mathrm{s}}\, \mathrm{RT})}\,,
\end{equation}
so a single slider, calibrated in seconds, imposes the same decay
rate on every line at any sample rate. Measured by Schroeder
integration~\cite{schroeder1965} on the compiled plugin, a setting of \SI{0.5}{\second}
yields a reverberation time of \SI{0.5000}{\second}
(\cref{fig:rt60}).

\section{Stability certificates}
\label{sec:certificate}

Before anything is shipped, the representation is analysed for
stability. The criterion is a small-gain argument~\cite{zames1966}.
For a loop of elements $E_1, \ldots, E_K$ in series, with
$E_k(e^{j\omega})$ the frequency response of element $k$ and
$\sigma_{\max}(\cdot)$ the largest singular value of a matrix, the
closed loop is stable if
\begin{equation}\label{eq:smallgain}
  \sup_{\omega}\; \prod_{k=1}^{K}
  \sigma_{\max}\!\bigl(E_k(e^{j\omega})\bigr) < 1\,,
\end{equation}
evaluated over a frequency grid, with pole checks on every recursive
filter section besides. The spectral radius is reported but is not
the criterion, since it says nothing once filters sit in the loop
and, for non-normal matrices, permits arbitrarily large transient
growth.

The analysis runs on the parameter values exactly as emitted, that
is, after decimal formatting and the single-precision cast of the
audio path. A trained orthogonal matrix in our evaluation measures
$\sigma_{\max} = 1.000000170$ after this chain, marginally above
unity, and an analysis of the pristine double-precision values would
therefore misdescribe the artefact that ships. A lossless prototype is
classified marginally stable; with the reverberation-time control
present, the loop is attenuated at every slider position and the
export is certified stable outright. The certificate is written as
JSON next to the generated code, and the plugin exporter refuses to
build a model whose verdict is unstable or unproven unless
explicitly overridden.

\section{Live training and deployment}
\label{sec:live}

In the demonstration an FDN of $N=4$ delay lines with learnable
output gains is optimised with Adam (learning rate \num{0.05}) for 200
steps at \SI{48}{\kilo\hertz}, minimising the mean-squared error (MSE)
between its frequency response and a target.
The training loop is made audible by a callback that re-emits the
model after each optimiser step and atomically rewrites a watched
\texttt{.dsp} file. A resident plugin built on the FAUST
interpreter~\cite{faust2clap} reloads the file when it changes,
recompiling the emitted network in under \SI{10}{\milli\second} on an
Apple M2, whilst ADAC itself re-emits the model in
\SI{0.2}{\milli\second}, so a step becomes audible within a few audio
buffers of the write. Slider values survive each reload by address, so
the reverberation-time control may be adjusted whilst the optimiser
works. Publishing is
deduplicated and rate-limited, so a converged or paused optimisation
causes no reloads. The host plugin is a CLAP effect~\cite{faust2clap}, wrapped as
VST3 and AU so that the same workflow runs in hosts without native
CLAP support.

Deployment is a single call comprising code generation,
certification, JUCE project generation, release compilation, and
installation of validated VST3 and AU binaries into the user's plugin
folders, with the macro-controls exposed as automatable parameters in
every format. From trained model to installed plugin is at most a few
minutes of compilation.

Because the generated code is standard FAUST, it can be pushed to
fixed-point targets through the Syfala flow~\cite{syfala}; these requantise the
coefficients, and the certificate is recomputed at the target word
length.

\vspace{-5mm}
\section{Conclusions}
\label{sec:conclusion}

The case study is an FDN, but ADAC is not
restricted to one. Any audio graph built from series, parallel, and
recursive composition with parameterised leaves passes through the
same traversal and emission unchanged, and a new leaf type requires
only a new emitter. As a larger instance, a scattering delay network (SDN)
for a six-wall room compiles through the same path; its thirty
wall-to-wall delay lines, block-diagonal
scattering~\cite{schlecht_scattering}, and per-wall input and output
routing reproduce the FLAMO source to within the same single-precision agreement reported for the FDN. What
remains is breadth of
leaf coverage and a stable specification of the representation against
which other frameworks and tools can be written.

A sharper limit is the linear, time-invariant (LTI) setting itself. The
emitter extends readily to nonlinear and time-varying processors,
since FAUST expresses waveshapers, envelope followers, and modulation
directly, and a frontend for a framework that carries such modules,
for example dasp~\cite{dasp} or differentiable time-varying
filters~\cite{diffapf}, would bring them into the same representation.
The equivalence and stability arguments are tied to that linear
setting, so carrying the guarantees across alongside the emitter is
the natural next step. Source code is available on
GitHub.\footnote{Code: \url{https://github.com/cucuwritescode/adac}.
Documentation: \url{https://adac.readthedocs.io}.}

\section{Acknowledgments}
The authors thank Gloria Dal Santo for the FLAMO framework and the FAUST team at GRAME for the compiler infrastructure.

\bibliographystyle{IEEEtranDAFx}
\bibliography{DAFx26_tmpl}

\end{document}